\documentclass[a4paper, 12pt]{article}

\usepackage[english]{babel}
\usepackage{amssymb}
\usepackage{graphicx}

\begin{document}

\begin{center}
\begin{large}
\textbf{To the theory of quasi-phase-matched parametric
amplification in periodically-poled optical nonlinear crystals}
\end{large}
\end{center}

\vspace{0.5cm} \centerline{E.V. Makeev and A.S. Chirkin}

e-mails: makeev@newmail.ru,~~~chirkin@squeez.phys.msu.su

\vspace{0.5cm}
\begin{center}
\centerline{\emph{\small Physics Department, M. V. Lomonosov
Moscow State University, Vorob'evy Gory, Moscow 119992 Russia}}
\end{center}

\vspace{0.5cm}
\centerline{\textbf{Abstract}}

\vspace {0.5cm} Theory of the optical parametric amplification at
high-frequency pumping in crystals with a regular space modulation
of the sign of nonlinear coupling coefficient of interacting waves
is developed. By applying the matrix method, the theory is based
on a step-by-step approach. It is shown that, in the case where
the pumping intensity is less than some critical value, the
spatial dynamics of the signal intensity inside a separate layer
with the constant nonlinear coefficient has an oscillatory
behavior and the change of the signal intensity from layer to
layer is defined, in general, by the power function. The same law
is valid for the change of variance of signal's quadrature
components. At large number of layers, these dependences can be
reduced to the well-known ones for homogeneous nonlinear optical
crystals.\\


\section{Introduction}

Quasi-phase-matched interactions of light waves as well as
phase-matched interactions provide the realization of an effective
energy exchange between interacting waves (see, for example, the
reviews \cite{Byer, Chirkin Volkov Laptev Morozov}) and the
possibility to obtain nonclassical light \cite{Baldi Ashieri Nouh
Micheli Ostrowsky, Serkland Fejer Byer Yamamoto}.
Quasi-phase-matched interactions are implemented in crystals with
the periodic modulation of nonlinear susceptibility. These are
crystals with the regular domain structure or periodically-poled
nonlinear crystals (PPNC). In such crystals, the mismatch of wave
vectors $\Delta k$ of interacting waves is compensated by an
inverse-lattice vector $\kappa$ of nonlinear susceptibility's
modulation $$\kappa=2\pi m/\Lambda,$$ where $\Lambda$ is the
modulation period and $m$ is the quasi-phase-matching order.

The conventional approach for implementing parametric interactions
in PPNC consists in a choice of the domain-structure period
$\Lambda$, in view of the expression $$\Lambda=2\pi/|\Delta k|,$$
where $\Delta k$ is the phase mismatch between interacting waves.
The period $\Lambda$ is usually determined from the analysis of
the second-harmonic generation \cite{Byer, Chirkin Volkov Laptev
Morozov}. At the coherence length \mbox{$l_c=\Lambda/2$}, the
second-harmonic intensity reaches the maximum value and the phase
(related to the phase mismatch and equal to $\pi$) compensates the
change of the nonlinearity sign in the neighboring layer (domain)
due to the optical-axis inversion.

While nonlinear optical interactions are calculated analytically,
PPNC is replaced by a homogeneous crystal with an effective
nonlinear coefficient $\beta_{\rm eff}$ distinguished from a value
$\beta$ for the homogeneous crystal by the factor $2/\pi m$, so
that $$\beta_{\rm eff}=\frac{2}{\pi m}\,\beta.$$ However, such
approach to the optical-parametric-amplification process is not
always valued. As it is known, in a parametric process the
transition to a steady state of the optimum phase relation
essentially depends on the pumping intensity \cite{Ahmanov Djakov
Chirkin, Chirkin Yusupov 1981}. Recently, the spatial dynamics of
intensity and phase of the parametrically amplified wave in PPNC
was investigated in detail \cite{Makeev Chirkin}. In particular,
it was shown that at the transition from layer to layer the sign
of derivative in the phase relation is changed. The change of the
signal wave intensity in PPNC also differs from the homogeneous
crystal case. In the present paper, the simple ``step-by-step"
approach to the analysis of quasi-phase-matched optical parametric
amplification is presented. The knowledge of transmission factors
of the layers is the basis of this approach. For the definition of
transmission factors, the matrix approach elaborated in
\cite{Beskrovnyy Baldi} is used. For an arbitrary number of layers
with thickness equal to the coherence length, expressions for the
signal intensity and variance of quadrature components for an
initial random signal phase are derived.

The paper is organized as follows.

In Sec. 2 the equations for degenerate optical parametric
amplification in PPNC are presented in generic form using
undepleted pump approximation. Section 3 contains the solution of
differential equation for the optical parametric amplification in
PPNC in matrix form. In Sec. 4 the solution obtained is analyzed
for the case satisfying the condition of quasi-phase-matched
interactions. In Sec. 5 the possibility to generate the
quadrature-squeezed light is considered when classical
fluctuations are suppressed. The results obtained are summarized
in Sec. 6.

\section{Basic Equations}

The optical parametric amplification process in PPNC of a wave
with frequency $\omega$ in the pumping-wave field with frequency
$2\omega$ is determined by the following equations:
\begin{equation}\label{eq: full system}
  \left\{
    \begin{array}{l}
      \displaystyle\frac{dA}{dz}=-i\beta g(z) A_pA ^*\exp(-i\Delta kz), \\
      \displaystyle\frac{dA_p}{dz}=-i\beta g(z) A^2\exp(i\Delta kz),
    \end{array}
  \right.
\end{equation}
where $A$ and $A_p$ are the complex amplitudes of the signal and
pumping waves, respectively, $\beta$ is the modulus of nonlinear
wave coupling coefficient, \mbox{$\Delta k=k-2k_p$} is the phase
mismatch, and $g(z)$ is the periodic function equal either to $+1$
or to $-1$ at the thickness of a separate layer, being dependent
on the nonlinear-coefficient sign. In the general case, it is
impossible to solve the set of equations (\ref{eq: full system})
analytically.

Within the framework of undepleted pump approximation, the process
under consideration is described by the equation
\begin{equation}\label{eq: undepleted pump equation}
\frac{dA}{dz}=-i\beta g(z)A_pA^*\exp (-i\Delta kz),
\end{equation}
where $A_p$ is the constant value. Equation (\ref{eq: undepleted
pump equation}) makes it possible to get an analytical solution on
each layer for given $g(z)$. However, the solution for arbitrary
number of layers, cannot be cast into a suitable for the
application form. In this connection, to solve Eq.~(\ref{eq:
undepleted pump equation}), we use the matrix method, which allows
one to obtain the solution in a convenient form. Firstly we
rewrite Eq. (\ref{eq: undepleted pump equation}) introducing a
reduced length \mbox{$\zeta=z/L_{\rm nl}$}:
\begin{equation}\label{eq: normalizing undepleted pump equation}
  \begin{array}{l}
     \displaystyle \frac{dA}{d\zeta}=-ig(\zeta)A^*\exp(-i\delta\zeta), \\
     \displaystyle A(\zeta=0)=A_0,
  \end{array}
\end{equation}
where \mbox{$L_{\rm nl}=1/\beta|A_p|$} is the so-called nonlinear
length, \mbox{$\delta=\Delta k L_{\rm nl}$} is the normalized
phase mismatch, and $A_0$ is the signal amplitude at the PPNC
input.

\section{Matrix Differential Equation and Its Discrete
Solution}

Equations (\ref{eq: normalizing undepleted pump equation}) by
means of the substitution
\begin{equation}\label{eq: B}
  A=B\exp(-i\delta\zeta/2)
\end{equation}
are reduced to the following system of equations:
\begin{equation}\label{eq: B system}
  \begin{array}{l}
    \displaystyle\frac{dB}{d\zeta}=\frac{i\delta B}{2}-ig(\zeta)B^*, \\
    \displaystyle\frac{dB^*}{d\zeta}=-\frac{i\delta B^*}{2}+ig(\zeta)B,
  \end{array}
\end{equation}
which can be presented in matrix form
\begin{equation}\label{eq: B equation}
  \frac{dC}{d\zeta}=DC,
\end{equation}
with the initial values \mbox{$C=C_0$}. In Eq. (6) $C_0$ is the
value of the matrix $C$ at \mbox{$\zeta=0$}
\begin{equation}\label{eq: matrix C}
C=
\left(
    \begin{array}{c}
      B \\
      B^*
    \end{array}
  \right), \qquad
C_0=
\left(
    \begin{array}{c}
      B_0 \\
      B_0^*
    \end{array}
  \right),
\end{equation}
and $D$ is the matrix determined by the expression
\begin{equation}\label{eq: D}
D(\zeta)=
  \left(
    \begin{array}{cc}
      i\delta/2 & -ig(\zeta) \\
      ig(\zeta) & -i\delta/2
    \end{array}
  \right).
\end{equation}
The function $g(\zeta)$ takes the value either $+1$ or $-1$. One
introduces matrices $D_\pm$ associated with these values
\begin{equation}\label{eq: D +-}
D_\pm = \frac{i}{2}
  \left(
    \begin{array}{cc}
      \delta & \mp 2 \\
      \pm 2 & -\delta
    \end{array}
  \right).
\end{equation}

Let PPNC have layers with the length \mbox{$\Lambda/2$}. It is
just the case used for repolarization of a nonlinear crystal. Let
\mbox{$g(\zeta)=1$} in the first layer. Then the solution of Eq.
(\ref{eq: B equation}) reads
\begin{equation}\label{eq: C (zeta)}
C(\zeta)=\exp(D_+\zeta)C_0,
\end{equation}
where the exponential function of matrix $D_+\zeta$ is nothing
else as the transmission factor for the layer and it is introduced
as follows:
\begin{equation}\label{eq: exp (A)}
\exp(D_+\zeta)=\sum_{k=0}^{\infty}\frac{(D_+\zeta)^k}{k!}.
\end{equation}
In accordance with (\ref{eq: C (zeta)}), at the output of the
first layer, we have
\begin{equation}\label{eq: C (Lambda/2)}
C(\Lambda/2)=\exp(D_+\Lambda/2)C_0.
\end{equation}
In order to obtain the signal at the output of the second layer,
we need to suppose \mbox{$D=D_-$} and take into account the fact
that the output of the first layer is simultaneously the input of
the second layer. As a result, the signal at the output of the
second layer is determined by the expression
\begin{equation}\label{eq: C}
C(\Lambda)=\exp(D_-\Lambda/2)C(\Lambda/2)=\exp(D_-\Lambda/2)\exp(D_+\Lambda/2)C_0.
\end{equation}
For $N$ layer pairs, Eq. (13) reads as follows:
\begin{equation}\label{eq: CN}
C(N\Lambda)=\Big[\exp(D_-\Lambda/2)\exp(D_+\Lambda/2)\Big]^NC_0=Q^NC_0,
\end{equation}
where the matrix $Q$ is
$$Q=\exp(D_-\Lambda/2)\exp(D_+\Lambda/2).$$ Thus, the matrix $Q^N$
defines the signal transformation from the PPNC input to its
output. The solution constructed by such a way is based on the
assumption that PPNC has a discrete structure. Due to this, we
call such approach the step-by-step approach.

\section{Quasi-Phase-Matched Interaction}

Now we calculate the matrix $Q$ in the case where the
quasi-phase-matching condition is realized and the modulation
period \mbox{$\Lambda=2l_c$}. The exponential matrix can be
presented in a compact form
\begin{equation}\label{eq: exp (D +-)}
\exp(D_\pm\Lambda/2)=
  \left(
    \begin{array}{cc}
      c+i\delta s & \mp2is \\
      \pm2is & c-i\delta s
    \end{array}
  \right),
\end{equation}
where functions $c(\Lambda)$ and $s(\Lambda)$ read
\begin{equation}\label{eq: c*s}
  c(\Lambda)=\cos(\mu\Lambda/4), \qquad
  s(\Lambda)=\mu^{-1}\sin(\mu\Lambda/4),
\end{equation}
with $$\mu=\sqrt{\delta^2-4}=2\sqrt{I_{\rm cr}/I_p-1}.$$ In these
relations, $I_p=|A_p|^2$ is the pumping intensity, $I_{\rm
cr}=(\Delta k/2\beta)^2$ is the so-called critical intensity of
the pumping wave or the parametric-trapping intensity. To use PPNC
for the optical parametric amplification makes sense at
$I_p<I_{\rm cr}$. In this case, the signal wave inside a separate
layer oscillates. By multiplying the exponential matrices $D_\pm$,
the matrix $Q$ can obtained in the form
\begin{equation}\label{eq: Q}
Q=
  \left(
    \begin{array}{cc}
      -4s^2+(c+is\delta)^2 & 4s^2\delta \\
      4s^2\delta & -4s^2+(c-is\delta)^2
    \end{array}
  \right).
\end{equation}
Since \mbox{$\Lambda=2l_c=2\pi/|\delta|$}, the functions (\ref{eq:
c*s}) are
\begin{equation}\label{eq: c*s (lopt)}
c(l_c)=\cos\left(\frac{\pi}{2}\frac{\mu}{\delta}\right), \qquad
s(l_c)=\mu^{-1}\sin\left(\frac{\pi}{2}\frac{\mu}{\delta}\right).
\end{equation}
At $|\delta|\gg2$ ($I_{\rm cr}\gg I_p$),  $\mu/|\delta|\approx1$,
therefore, we arrive at
\begin{equation}\label{eq: Q(lopt)}
Q\approx -\frac{1}{\mu^2}
  \left(
    \begin{array}{cc}
      -4-\delta^2 & 4 |\delta| \\
      4|\delta| & -4-\delta^2
    \end{array}
  \right).
\end{equation}
Raising the matrix (\ref{eq: Q(lopt)}) to $N$th power, we obtain
\begin{equation}\label{eq: Q(lopt)^N}
Q^N=\frac{1}{2}
  \left(
    \begin{array}{cc}
      \alpha^{-N}+\alpha^N & \alpha^{-N}-\alpha^N \\
      \alpha^{-N}-\alpha^N & \alpha^{-N}+\alpha^N
    \end{array}
  \right), \qquad
  \alpha=\left(\frac{|\delta|+2}{|\delta|-2}\right).
\end{equation}
Taking into account expressions (\ref{eq: matrix C}) and (\ref{eq:
CN}), we arrive at the following result for $B$:
\begin{equation}\label{eq: Blopt}
B=\frac{1}{2}\Big[(\alpha^{-N}+\alpha^N)B_0+(\alpha^{-N}-\alpha^N)B_0^*\Big].
\end{equation}
Now in view of the relations $$A_0=a_0e^{i\varphi_0}\qquad\mbox{
and}\qquad I_0=|A_0|^2,$$  we obtain the following expression for
the signal wave intensity $I_N=|B|^2=|A|^2$ at the PPNC output:
\begin{equation}\label{eq: INloptphi0}
  I_N=\Big[\alpha^{-2N}\cos^2\varphi_0+\alpha^{2N}\sin^2\varphi_0\Big]I_0.
\end{equation}
One can see that it is just the well-known property of phase
sensitivity of degenerate parametric amplification. From Eq.
(\ref{eq: INloptphi0}) follows that the largest amplification
takes place at the signal phase \mbox{$\varphi_0=\pm(\pi/2)$}. The
largest signal suppression is achieved at the signal phase
\mbox{$\varphi_0=\pm\pi$}. For number of layer pairs
\mbox{$N\gg1$}, at \mbox{$|\delta|\gg2$} expression (\ref{eq:
INloptphi0}) can be transformed into the following one:
\begin{equation}\label{eq: INexp}
  I_{hc}=\Big[\cosh (\Gamma L)-\cos(2\varphi_0)\sinh (\Gamma L)\Big]I_0,
\end{equation}
where $\Gamma=(4/\pi)\beta|A_p|$ is the parametric amplification
increment and \mbox{$L=2Nl_c$} is the total length of PPNC. It is
worthy to compare expression (\ref{eq: INexp}) with the analogous
one for the case of a homogeneous crystal under the phase-matching
condition. In the latter case, the parametric amplification
increment is equal to $\Gamma_0=2\beta_0|A_p|$ (see, for example,
\cite{Ahmanov Djakov Chirkin}), where $\beta_0$ is the nonlinear
wave coupling coefficient for the homogeneous crystal. The ratio
$\Gamma/\Gamma_0= (2/\pi)\beta/\beta_0$ shows the difference
between the optical parametric amplification in PPNC and in the
homogeneous crystal. In the case $\beta=\beta_0$, we obtain the
well-known result $\Gamma/\Gamma_0=2/\pi$.

Thus, at the quasi-phase-matched-wave interaction, the intensity
of parametrically amplified signal wave under the conditions
mentioned above varies in the same way as in the homogeneous
crystal with the effective nonlinear coefficient \mbox{$\beta_{\rm
eff}=(2/\pi)\beta$}. It should be noted that the same value
$\beta_{\rm eff}$ for PPNC can be obtained, if truncated equation
(\ref{eq: undepleted pump equation}) is averaged over the
modulation period of nonlinear-wave-coupling coefficient. However,
in the latter case the conditions of applicability of the
equations obtained are not evident.

Using formula (\ref{eq: INloptphi0}) we have calculated curves
presented in Figs.~1 and 2 for $\varphi_0=\pi/2$. Figure~1 shows
the dependence of the signal intensity on the interaction length
at the quasi-phase-matched parametric amplification in PPNC.
Figure~2 demonstrates the dependence of the parametric
amplification increment on the interaction length at different
relations between the pumping intensity and its critical value.
From these figures, one can see that, at the same pumping
intensity and given interaction length, the signal wave intensity
increases when the pumping intensity approaches its critical
value. Therefore, the replacement of PPNC by the homogeneous
crystal with the effective nonlinear coefficient provides less
value for the signal intensity. This difference grows up when the
pumping intensity $I_p$ approaches to $I_{\rm cr}$. Comparison of
the conversion efficiency for the parametric amplification in PPNC
and in the homogeneous crystal with nonlinear coefficient
$\beta_{\rm eff}$ is done in Fig.~3. Note that at $I_p\simeq
I_{\rm cr}$ we have used general formula (\ref{eq: Q}).

\section{Quadrature Components}

We turn now to the analysis of quadrature components $X$, $Y$ of
the signal wave in PPNC:
\begin{equation}\label{eq: X*Y}
X(z)=\frac{1}{2}\left[A(z)+A^*(z)\right], \qquad
Y(z)=\frac{1}{2i}\left[A(z)-A^*(z)\right].
\end{equation}
According to Eq. (\ref{eq: Blopt}), one has for $X$-quadrature
\begin{equation}\label{eq: X}
X(z)=\frac{1}{2}\left[(\alpha^{-N}+\alpha^N)\cos\phi_-
+(\alpha^{-N}-\alpha^N)\cos\phi_+\right],
\end{equation}
where the phase $\phi_\pm=\varphi_0\pm\delta\zeta/2$.

To obtain a signal with suppressed fluctuations of one of the
quadrature components, the random signal with a uniform
distribution function for the initial phase $\varphi_0$ is worthy
to consider
\begin{equation}\label{eq: w}
  w(\varphi_0)=\frac{1}{2\pi}, \qquad -\pi\leq\varphi_0\leq\pi.
\end{equation}
Then the mean value $\langle X(z)\rangle=0$ and the variance of
quadrature component \mbox{$\langle
X^2(z)\rangle=\overline{X^2(z)}-(\overline{X}(z))^2$} is given by
the expression
\begin{equation}\label{eq: <X^2>}
  \langle X^2(z)\rangle=\frac{1}{2}a_0^2\left[\alpha^{2N}+
  (\alpha^{-2N}-e^{2N})\cos^2(\delta\zeta/2)\right].
\end{equation}
At $\delta\zeta=\Delta k L=2\pi n$ ($n=\pm1, \pm2, \ldots$),
$X$-quadrature fluctuations are reduced, namely,
\begin{equation}\label{eq: <X^2> min}
  \langle X^2(z)\rangle_{\rm min}=\frac{1}{2}a_0^2\alpha^{-2N}.
\end{equation}
In the case $\delta\zeta=\pi(1+2n)$, the variance reaches the
maximum value:
\begin{equation}\label{eq: <X^2> max}
  \langle X^2(z)\rangle_{\rm max}=\frac{1}{2}a_0^2\alpha^{2N},
\end{equation}
i.e., fluctuations of quadrature components are amplified.

Thus, in PPNC the spatial changing of the variance of signal's
quadrature components is determined by the power functions as well
as the signal intensity. However, if number of layer pairs
$N\gg1$, expressions (\ref{eq: <X^2>})--(\ref{eq: <X^2> max}) can
be transformed into the forms similar to the ones for the
homogeneous crystal:
\begin{equation}\label{eq: <X^2> min max}
  \langle X^2(z)\rangle_{\rm min/max}=\langle X^2(0)\rangle e^{\mp\Gamma L}.
\end{equation}
It is not difficult to draw a conclusion, in view of formulas
(\ref{eq: <X^2> min}) and (\ref{eq: <X^2> min max}), that
suppression of fluctuations in PPNC occurs more effectively than
in the homogeneous optical crystal with the effective nonlinear
coefficient.

\section{Conclusions}

The application of step-by-step matrix approach allows one to
calculate the optical parametric amplification in
periodically-poled nonlinear crystals without analysis of the
spatial dynamics inside a separate layer. The method can be
applied to analysis of the processes in PPNC at an arbitrary
period of the crystal repolarization and arbitrary relation of the
critical intensity to the pumping intensity. In other words, the
method is not restricted to the conditions $I_p\ll I_{\rm cr}$ and
$N\gg1$ which are necessary for the correct analysis in the
traditional-approach case where PPNC is replaced by an homogeneous
crystal with effective nonlinear coefficient. This circumstance
allows one to use this method for both analytical and numerical
\cite{Beskrovnyy Baldi} calculations.

The spatial dynamics of the signal intensity and quadrature
components was analyzed versus the ratio $I_p/I_{\rm cr}$ and the
interaction length. It was shown that at the pumping intensity
less than the critical value, the spatial distribution of the
signal intensity inside a separate layer exhibits an oscillatory
character and changing of the signal intensity and variance of
quadrature components from layer to layer follow the power
function dependences. In the case of larger number of layers and
$I_p\ll I_{\rm cr}$, these dependences can be reduced to
well-known formulas for homogeneous nonlinear crystals. Notice
that the requirement $I_p\geq I_{cr}$ corresponds to the
parametric trapping in the homogeneous crystal.

It should be emphasized that our analyze has shown that the
quasi-phase-matched optical parametric process in PPNC occurs,
generally speaking, more effectively than the parametric process
in the homogeneous crystal with the nonlinear coupling coefficient
equal to the effective nonlinear coefficient for PPNC. The
generalization of the approach presented here onto the case of
quasi-phase-matched consecutive wave interactions and
nonstationary parametric interactions is in progress.

\section*{Acknowledgments}

 The authors are grateful to I. V. Golovnin, G. D. Laptev,
E. Y. Morozov, and A. A. Novikov for helpful discussions.

The work was partially supported by INTAS under
Project~No.~01-2097.

\newpage
\begin{figure}[h]
%\vspace{0cm} \hspace{0cm}
\centerline{\includegraphics[width=.8\textwidth]{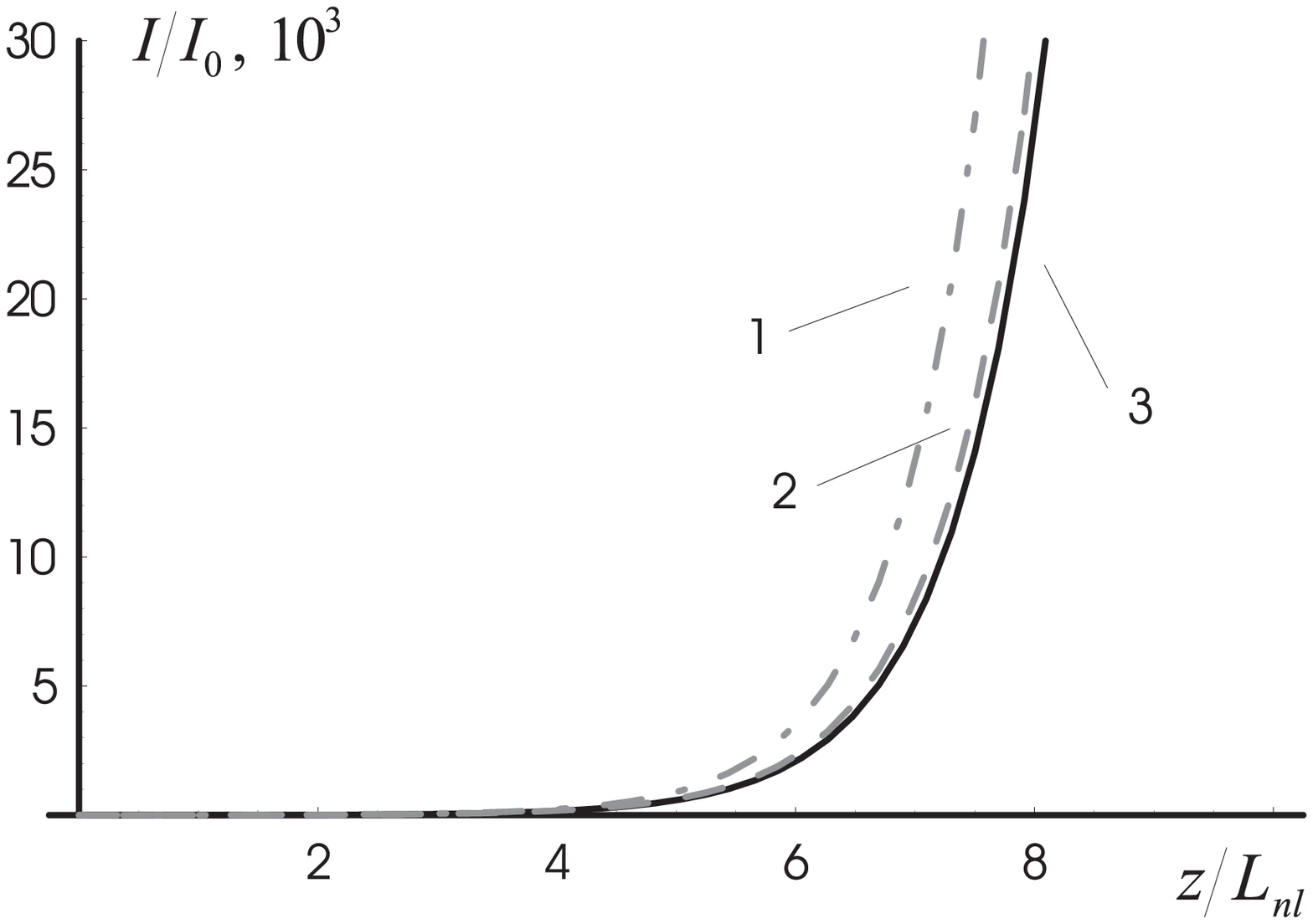}}
\caption{\small Coefficient of the parametric signal amplification
in PPNC as a function of the interaction length for the phase
$\varphi_0=\pi/2$ and different ratios of the pumping intensity
$I_p$ to the critical intensity $I_{\rm cr}$: $I_p=I_{\rm
cr}$~(1), $I_p<I_{\rm cr}$~(2), and $I_p\ll I_{\rm cr}$~(3).
Curve~2 corresponds to formula (\ref{eq: INloptphi0}) and curve~3
corresponds to formula (\ref{eq: INexp}).}\label{fig: 2D}
\end{figure}

\newpage
\begin{figure}[h]
%\vspace{0cm} \hspace{0cm}
\centerline{\includegraphics[width=.8\textwidth]{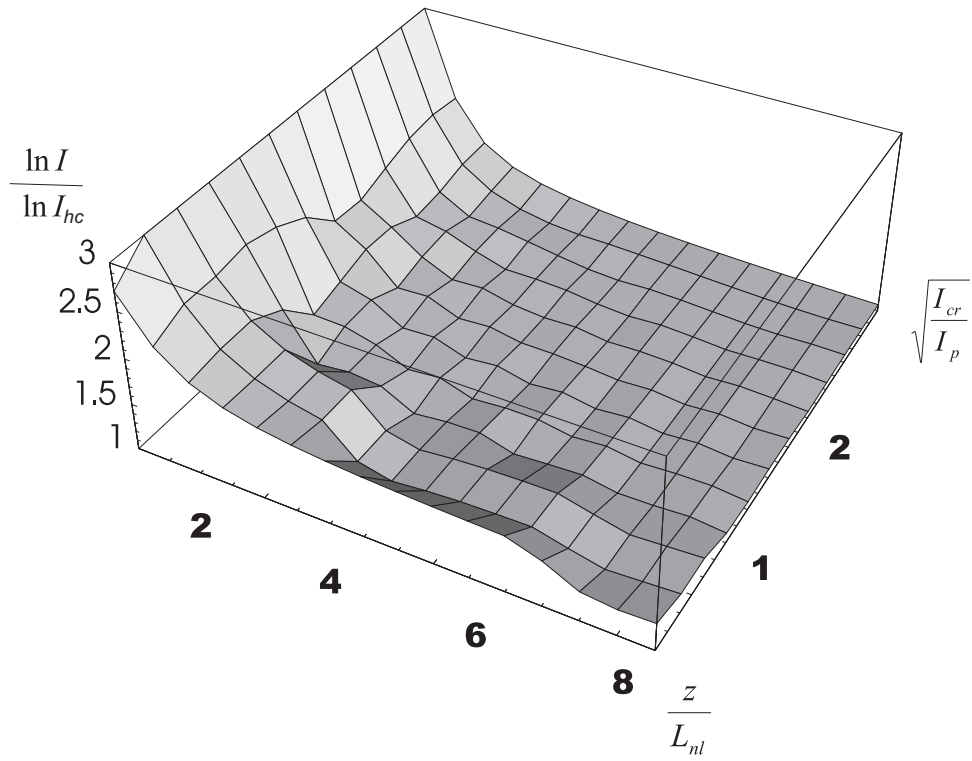}}
\caption{\small Relative increment of the parametric amplification
in PPNC as a function of the interaction length and ratio $I_{\rm
cr}/I_p$. $I$ and $I_{\rm hc}$ are the signal intensities for the
case of PPNC and homogeneous nonlinear crystal,
respectively.}\label{fig: 3D}
\end{figure}

\newpage
\begin{figure}[h]
%\vspace{0cm} \hspace{0cm}
\centerline{\includegraphics[width=.8\textwidth]{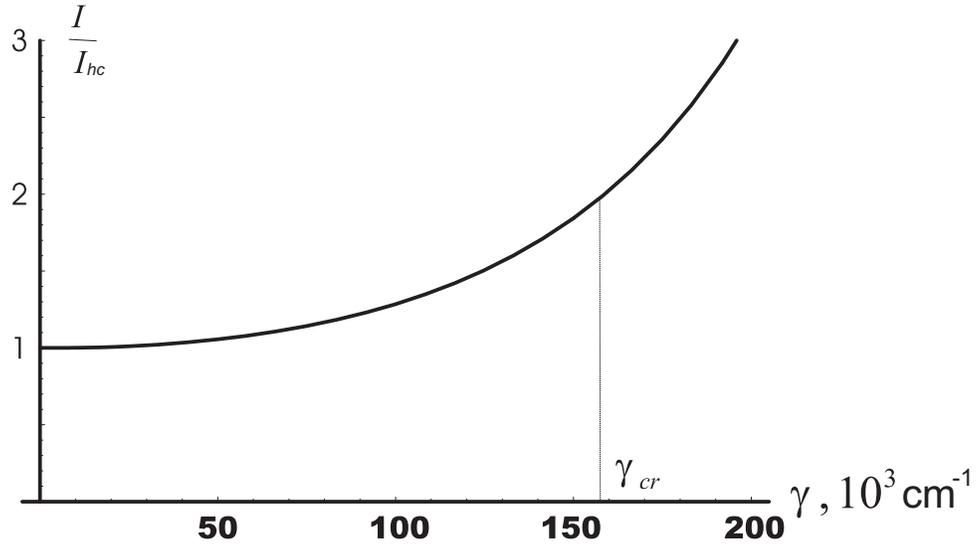}}
\caption{\small Ratio of the parametrically-amplified signal
intensity $I$ in a LiNbO$_3$ PPNC to the intensity $I_{\rm hc}$ in
a LiNbO$_3$ homogeneous crystal with corresponding effective
nonlinear coefficient $\beta_{\rm eff}$ versus
$\gamma=\beta\sqrt{I_p}$ for the $eee$-interaction type and
$\lambda=0.5~\mu $m, $\Lambda=20~\mu $m, $N=3$, and $\gamma_{\rm
cr}=\beta\sqrt{I_{\rm cr}}$.}\label{fig: 1D}
\end{figure}

\end{document}